\documentclass[aps,prb, two column  , superscriptaddress,showpacs,amsmath,amssymb]{revtex4}

\usepackage[dvips]{graphicx}
\usepackage{color}
\definecolor{DarkBlue}{rgb}{0.1,0.1,0.5}
\definecolor{Red}{rgb}{0.9,0.0,0.1}
\definecolor{Green}{rgb}{0.0,0.99,0.0}

\newcommand{\bra}[1]{\left<#1\right\vert}
\newcommand{\ket}[1]{\left\vert#1\right>}

\newcommand{\threej}[6]{\left(\begin{array}{ccc}#1&#2&#3\\#4&#5&#6\end{array}\right)}

\begin{document}

\title{Resonant Enhancement of Charge Density Wave Diffraction in the Rare-Earth Tri-Tellurides}

\author{W. S. Lee}
\affiliation{Stanford Institute for Materials and Energy Sciences, SLAC National Accelerator Laboratory, Menlo Park, CA 94025, USA}

\author{A.~P.~Sorini}
\affiliation{Stanford Institute for Materials and Energy Sciences, SLAC National Accelerator Laboratory, Menlo Park, CA 94025, USA}
\affiliation{Lawrence Livermore National Laboratory, Livermore, CA 94550, USA}

\author{M. Yi}
\affiliation{Stanford Institute for Materials and Energy Sciences, SLAC National Accelerator Laboratory, Menlo Park, CA 94025, USA}
\affiliation{Department of Applied Physics, Stanford University, Stanford, CA 94305, USA}

\author{Y. D. Chuang}
\affiliation{Advanced Light Source, Lawrence Berkeley National Laboratory, Berkeley, CA 94720, USA}

\author{B. Moritz}
\affiliation{SIMES, SLAC National Accelerator Laboratory, Menlo Park, CA 94025, USA}

\author{W. L. Yang}
\affiliation{Advanced Light Source, Lawrence Berkeley National Laboratory, Berkeley, CA 94720, USA}

\author{J. -H. Chu}
\affiliation{Department of Applied Physics, Stanford University, Stanford, CA 94305, USA}

\author{H. H. Kuo}
\affiliation{Department of Applied Physics, Stanford University, Stanford, CA 94305, USA}

\author{A. G. Cruz Gonzalez}
\affiliation{Advanced Light Source, Lawrence Berkeley National Laboratory, Berkeley, CA 94720, USA}

\author{I. R. Fisher}
\affiliation{Stanford Institute for Materials and Energy Sciences, SLAC National Accelerator Laboratory, Menlo Park, CA 94025, USA}
\affiliation{Department of Applied Physics, Stanford University, Stanford, CA 94305, USA}

\author{Z. Hussain}
\affiliation{Advanced Light Source, Lawrence Berkeley National Laboratory, Berkeley, CA 94720, USA}

\author{T. P. Devereaux}
\affiliation{SIMES, SLAC National Accelerator Laboratory, Menlo Park, CA 94025, USA}

\author{Z. X. Shen}
\affiliation{Stanford Institute for Materials and Energy Sciences, SLAC National Accelerator Laboratory, Menlo Park, CA 94025, USA}
\affiliation{Department of Applied Physics, Stanford University, Stanford, CA 94305, USA}


\date{\today}

\begin{abstract}
  We performed resonant soft X-ray diffraction on known charge density wave (CDW) compounds, rare earth tri-tellurides. Near the $M_5$ ($3d$ - $4f$) absorption edge of rare earth ions, an intense diffraction peak is detected at a wavevector identical to that of CDW state hosted on Te$_2$ planes, indicating a CDW-induced modulation on the rare earth ions. Surprisingly, the temperature dependence of the diffraction peak intensity demonstrates an exponential increase at low temperatures, vastly different than that of the CDW order parameter. Assuming $4f$ multiplet splitting due to the CDW states,we present a model to calculate X-ray absorption spectrum and resonant profile of the diffraction peak, agreeing well with experimental observations. Our results demonstrate a situation where the temperature dependence of resonant X-ray diffraction peak intensity is not directly related to the intrinsic behavior of the order parameter associated with the electronic order, but is dominated by the thermal occupancy of the valence states.

\end{abstract}

\pacs{Valid PACS appear here}
\maketitle

\section {Introduction}
Electronic orders, such as charge, spin, and orbital orders, are important ground states that have been found in many correlated electron systems. These electronic orders are periodic real space modulations of a fraction of valence electrons, which can be uniquely probed via resonant X-ray diffraction. \cite{Abbamonte02, Abbamonte04,Christian05, Dhesi92,Hill95,Thomas04,Wilkins03} In the measurement of resonant X-ray diffraction, the incident photon energy is varied across an absorption edge of a constituent element of the sample, which allows virtual transitions of a core electron into unoccupied intermediate valence states associated with the absorption process. These processes result in a significant enhancement of the scattering cross section from the valence electrons, revealing the nature of valence states that are associated with electronic order. In most cases, the temperature dependence of the resonant diffraction peak exhibits a good correspondence with that of an order parameter. However, due to the complication of the resonance processes involved, there are circumstances when the resonant diffraction peak intensity may not be governed by the intrinsic temperature dependence of the order parameters.

\begin{figure} [t]
\includegraphics [clip, width=3.25 in]{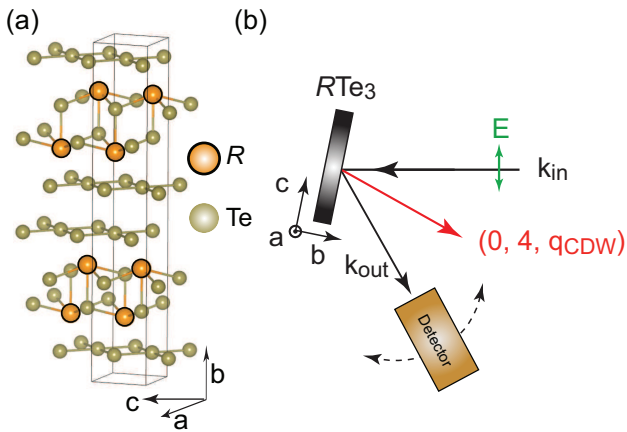}
\caption{\label{Fig1:ExperimentGeometry} (Color online) (a) Crystal structure of the rare earth tri-telluride, $R$Te$_3$ (\emph{R}=Tb, Dy, etc). (b) Scattering geometry. The polarization of the photons is set to be parallel to the scattering plane ($\pi$ geometry).
}
\end{figure}

We demonstrate such an effect in a family of rare earth tri-tellurides, \emph{R}Te$_3$ (\emph{R} is an rare earth ion, such as Tb, etc.), which is known to exhibit charge density wave (CDW) states. The crystal structure of \emph{R}Te$_3$ is shown in Fig. \ref{Fig1:ExperimentGeometry} (a); it consists of $R$Te and Te$_2$ layers stacked along the \emph{b}-axis. It has been shown that electrons in the Te \emph{p}-orbitals in the Te$_2$ sheets make up the Fermi surface, which is gapped at the portion of Fermi surface connected by the CDW wavevector. This suggests that the CDW state in $R$Te$_3$ is driven by the mechanism of Fermi surface nesting \cite{Gweon98, Brouet04, Brouet08, Moore08}. In addition, The CDW transition induces a lattice distortion which can be detected by non-resonant X-ray diffraction experiment \cite{Nancy08}. The rare earth ions, which are located in the adjacent layers, can be affected by the formation of the CDW state on the Te$_2$ plane. In this paper, we present the results of resonant X-ray diffraction measurements near the rare earth $M_5$ ($3d$-$4f$) absorption edge to study the CDW-induced effect on the rare earth ions, whose temperature dependence is, however, distinct from the expected CDW order parameter.

This paper is organized as follows. Materials and experimental method is described in Section II. Experimental results are demonstrated in Section III. Theoretical model and calculations are detailed in Section IV. Finally, a discussion is given in Section V.

\section{Materials and Experimental Methods }
Single crystals of TbTe$_3$, DyTe$_3$, CeTe$_3$, GdTe$_3$, and LaTe$_3$ were chosen for this study. At room temperature all of these compounds exhibit a unidirectional CDW with wavevector q $\sim$ 2/7 c*. DyTe$_3$ \cite{Nancy06} and TbTe$_3$ \cite{Banerjee} also exhibit a second CDW transition below 50 K with a wavevector oriented along the a-axis direction. In this study, we only focus on the CDW state occurring at higher transition temperatures. These crystals were grown by slow cooling a binary melt, as described previously\cite{Nancy06}. The resonant X-ray scattering experiments were performed at Beamline 8 of the Advanced Light Source (ALS) at Lawrence Berkeley National Laboratory using a two-circle diffractometer. The scattering geometry is sketched in Fig. \ref{Fig1:ExperimentGeometry} (b). The scattering plane is in the b-c plane with the polarization of the incident photon in the scattering plane ($\pi$ geometry). In order to obtain a clean and non-oxidized surface, the samples were cleaved in the air immediately prior to loading into the vacuum chamber. During the measurement, the samples were mounted on a 6-axis sample stage which is thermally connected to a low temperature cryostat with constant flow of liquid helium. The sample temperature was controllably varied from 10 K to 300 K with a fluctuation less than 1 K.

\section{Experimental Results}
\begin{figure} [t]
\includegraphics [clip, width=2.5 in]{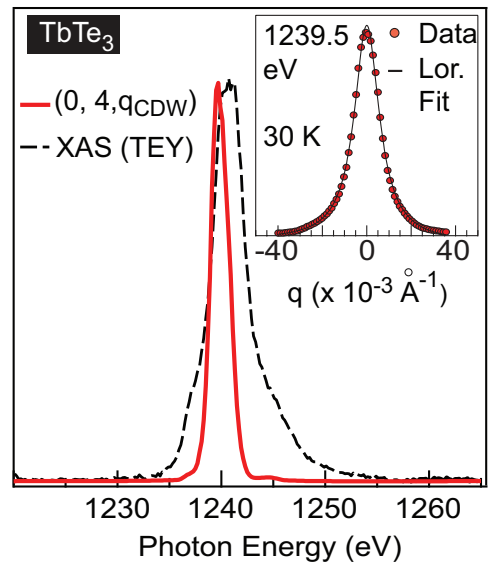}
\caption{\label{Fig2:RP_TbTe3} (Color online) The absorption curve (black) near the Tb $M_5$-edge and the intensity of the (0, 4, q$_{CDW}$) peak as a function of the incident photon energy, where q$_{CDW}$ $\approx$ 0.296 c*. Inset shows a representing $\theta$-2$\theta$ scan of the (0, 4, q$_{CDW}$) diffraction peak.
Data were taken at a temperature of 30 K.}
\end{figure}

We first focus on the results of TbTe$_3$, whose transition temperature to the CDW state (T$_{CDW}$) is known to be 335 K \cite{Nancy08}. At 30 K (inset of Fig. \ref{Fig2:RP_TbTe3}), CDW diffraction peak located at (0, 4, q$_{CDW}$ $\sim$ 0.296 c*) can be clearly resolved near the Tb $M_5$ edge (3$d$ - 4$f$ transition). This confirms that some sort of periodic modulation does occur on the TbTe rare earth layer, which is clearly induced by the formation of the CDW state in the neighboring Te$_2$ planes. Interestingly, as demonstrated in Fig. \ref{Fig2:RP_TbTe3}, the photon energy dependence of the diffraction peak intensity exhibits a lineshape that is different from the X-ray absorption spectrum (XAS). More specifically, the diffraction peak intensity is found to be strongest not when the incident photon energy is tuned at the maximum of the XAS, but 1eV below it, at 1239.5eV.  Once the photon energy is tuned away from 1239.5 eV, the peak intensity drops significantly, indicating that the resonant effect is significantly weakened, even though the photon energy is still within the absorption edge. It is found that the diffraction peak intensity is enhanced by approximately three orders of magnitude compared to that measured off-resonance at 1230 eV.  The distinct lineshape between the X-ray absorption curve and the resonant profile, defined as the diffraction peak intensity as a function of incident photon energy (e.g. red curve in Fig. \ref{Fig2:RP_TbTe3}), implies an electronic nature of the diffraction peak intensity, rather than the scattering by periodic lattice distortion, whose resonant profile would follow the main features of the XAS \cite{Abbamonte06}. As will become clear in Section IV, this resonant profile can be explained by the periodic $4f$ multiplet splitting with an identical period as that of the CDW state in the Te$_2$ plane.

\begin{figure*}[ht]
\includegraphics [clip, width=6.0 in ]{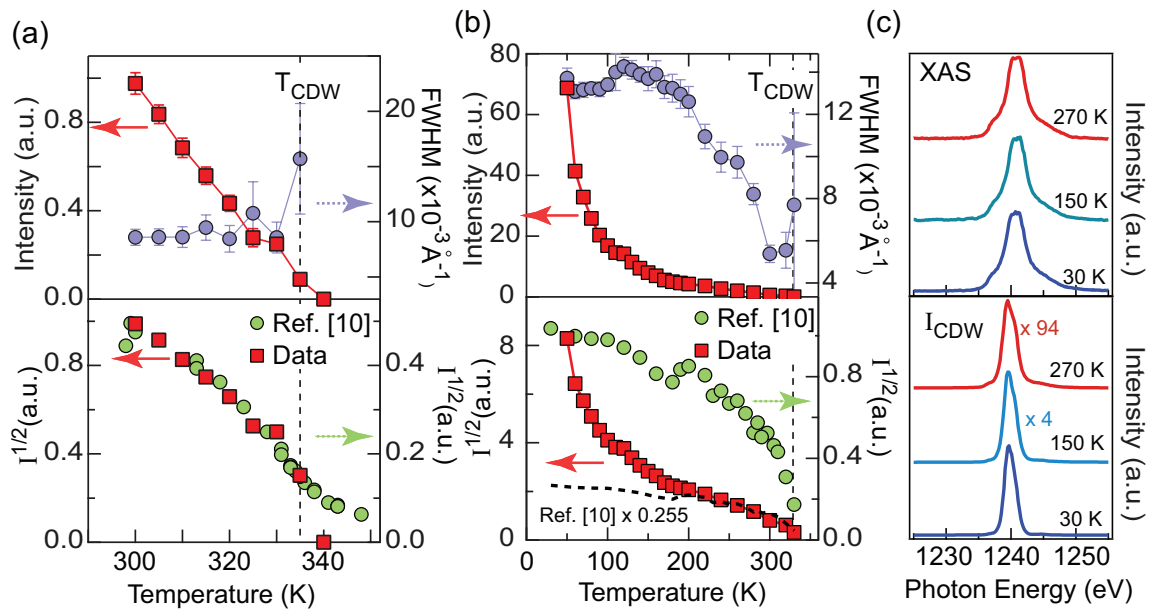}
\centering
\caption{\label{Fig3:T_dep_TbTe3} (Color online) (a) Upper panel: the intensity (red) and the width (blue) of the diffraction peak as a function of temperatures taken at incident photon energy of 1239.5 eV. Lower panel: the square root of the diffraction peak intensity, for the CDW diffraction peak obtained from both resonant soft X-ray diffraction (this experiment, red markers) and non-resonant hard X-ray diffraction (Ref. \cite{Nancy08}, green). (b) Same as (a), but the data taken at lower temperatures are plotted. (c) XAS (upper) and CDW resonance profiles (lower) at three representing temperatures. The structure of XAS and CDW resonant profile are essential unchanged.
}
\end{figure*}

The temperature dependence of the CDW resonant diffraction peak taken at 1239.5 eV at temperatures near T$_{CDW}$ is plotted in Fig. \ref{Fig3:T_dep_TbTe3} (a). The intensity of the peak decreases linearly with increasing temperature, and disappears when the sample temperature is higher than T$_{CDW}$. In addition, the width of the diffraction peak becomes broader at temperature near T$_{CDW}$. The order parameter, defined as the square root of the peak intensity, is shown in the lower panel of Fig. \ref{Fig3:T_dep_TbTe3}(a) and superimposed with the results obtained from non-resonant hard X-ray diffraction measurements. Near T$_{CDW}$, we found a good agreement on the temperature dependence of the diffraction peak between the measurement at the absorption edge of Tb $M_5$-edge and far away from any of the absorption edge, behaving like an order parameter in a 2nd order phase transition.

Surprisingly, upon cooling the sample to temperatures well below T$_{CDW}$, the temperature dependence of this CDW resonant diffraction peak deviates from the expected behavior of a CDW order parameter. As shown in Fig. \ref{Fig3:T_dep_TbTe3}(b), the intensity of the peak dramatically increases at temperatures below 200 K. In addition, the temperature dependence of the width is also unusual: the width first decreases below T$_{CDW}$; it then increases at temperatures lower than 300 K, until the temperature reaches 150 K - 200 K, where the width is seen to keep at a constant value.  Clearly, such anomalous resonant diffraction peak intensity enhancement at low temperatures does not reflect the magnitude of the CDW order parameter. This is further demonstrated in the lower panel of Fig. \ref{Fig3:T_dep_TbTe3}(b), the square root of the non-resonant hard X-ray diffraction peak intensity, which is proportional to the order parameter from lattice distortions, asymptotically approaches a constant value at low temperatures, in contrast to the temperature dependence of the resonant diffraction peak. Furthermore, according to angle-resolved photoemission measurement \cite{Brouet04, Brouet08}, the temperature dependence of the CDW gap, which reflects the CDW order parameters directly from the single particle spectral function, also exhibits a saturating behavior at low temperatures, as expected from a typical 2nd order phase transition. Therefore, the observed anomalous temperature dependence of the CDW resonant diffraction peak at Tb $M_5$ edge is related to the properties of the Tb 4$f$ states involved in the resonant process, in addition to the actual CDW order parameter.

\begin{figure*}[ht]
\includegraphics [clip, width=6.0 in]{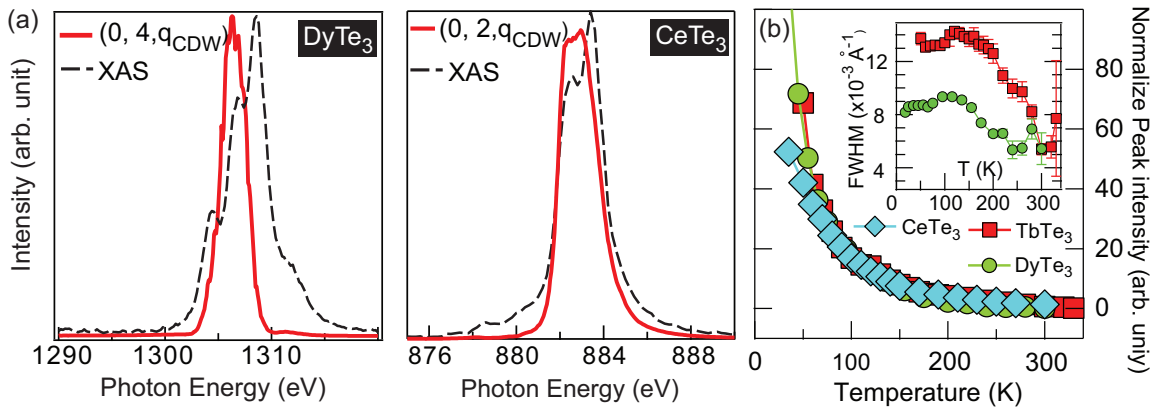}
\centering
\caption{\label{Fig4:OtherRTe3} (Color online) (a) the XAS (black) and the CDW resonant profile(red) of DyTe$_3$ (left) and CeTe$_3$ (right). Please note that the CDW diffraction peak for CeTe$_3$ is (0, 2, q$_{CDW}$) due to the smaller momentum transfer of the photons at Ce $M_5$-edge. (b) The CDW diffraction peak intensity of the TbTe$_3$, DyTe$_3$ , and CeTe$_3$ are plotted and superimposed. Inset  demonstrates the width of the diffraction peaks of TbTe$_3$ and DyTe$_3$.
}
\end{figure*}

In the lower panel of Fig. \ref{Fig3:T_dep_TbTe3}(c), the resonant profiles of this CDW diffraction peak at three representative temperatures are plotted. As can be seen, the lineshape of the resonant profile does not change as a function of temperature, while the overall intensity changes significantly. This confirms that the anomalous temperature dependence of the resonant diffraction peak shown in Fig.~\ref{Fig3:T_dep_TbTe3}(a) is not due to the change of its resonance profile. Furthermore, we do not resolve noticeable temperature dependence in XAS, as also shown in Fig.~\ref{Fig3:T_dep_TbTe3}(c). This suggests that the underlying mechanism for the temperature dependence anomaly is a subtle effect on the Tb 4$f$ states that can only be detected by a sensitive experimental probe such as resonant X-ray diffraction through the interference effect between rare earth ions with small variations in population of the crystal-field-split $4f$ $J$-multiplet.

The aforementioned behavior of the resonant diffraction peak is quite generic in the family of rare earth tri-tellurides. In Fig.~\ref{Fig4:OtherRTe3}(a) and (b), we plot XAS and resonant profile for DyTe$_3$ and CeTe$_3$ taken near the Dy and Ce $M_5$ edges, respectively. Similar to the results of TbTe$_3$, the resonant profile of the CDW diffraction peak near the rare earth $M_5$-edge is found to be different and narrower than the XAS, suggesting the electronic character of the resonant diffraction peak. The temperature dependence of the peak intensity and width for TbTe$_3$, DyTe$_3$, and CeTe$_3$ are plotted in Fig.~\ref{Fig4:OtherRTe3} (c) as a comparison. Intriguingly, similar to TbTe$_3$, the exponential-like temperature dependence of the resonant diffraction peak is also seen for both DyTe$_3$ and CeTe$_3$. This further indicates that the anomalous temperature dependence of the CDW resonant diffraction peak is related to a common nature of the rare earth 4$f$ state under the influence of CDW states.

We note that we have also measured GdTe$_3$ and LaTe$_3$; however, the intensity of the CDW diffraction peak is too weak to be detected, which provide another support that the CDW diffraction at \emph{M}-edge seen in other tri-telluride is not due to the scattering of the underlying lattice distortion due to CDW formation, but is due to electronic origins. The absence of any resonant effect at the Gd$^{3+}$ and La$^{3+}$ $M_5$-edge provide additional information about the nature of the resonant diffraction process at the rare earth $M_5$-edge. In particular, because neither Gd$^{3+}$ ($4f^7$) nor La$^{3+}$ ($4f^0$) have a spatially degenerate ground state (both are $L=0$ states by Hund's rules), the symmetry breaking due to the Te$_2$ plane CDW does not affect these ions and no signal is observed. These will be discussed in detail in the next section.

\section{Theoretical Calculation}
In order to explain the observed resonant profile and the temperature dependence, we calculated XAS for rare earth $R^{3+}$ ions in $R$Te$_3$ compounds at the rare earth $M_5$ edge. We also calculated the resonant elastic X-ray scattering (REXS) intensity for scattering from a model charge density wave (CDW) and compare this with the observed resonant diffraction profile. In our model, the non-zero resonant CDW signal at rare earth $M_5$ edge exists because the CDW modifies the crystal-field environment experienced by the $4f$ ions, affecting the crystal electric field (CEF) splitting of the rare earth ion's $J$-multiplet. Thermal population of the CEF eigenstates leads to the observed temperature dependence of the resonant intensity. In principle the local symmetry of the rare earth ions can be determined from knowledge of the atomic displacements in the CDW state. However, in order to yield a tractable calculation of the REXS spectra we develop a simpler model in which the point symmetry of the rare earth ions is reduced from O$_3$ to O$_2$ with the CDW periodicity and direction.

For our $R^{3+}$ REXS calculations, we treat the frozen CDW as producing a spatially modulated electrostatic field at the $R^{3+}$ site. To simplify the calculation, we assume that there are just two types of $R^{3+}$ sites: (a) sites at which the effective electrostatic field is maximal (we call these sites ``type-1");(b) sites at which the effective electrostatic field is weak or zero (``type-2"). The CDW periodicity is approximated by a real space pattern of type 1-2-1. The absorption coefficients for a type-1 site, $\mu_1$, would be  different from those of type-2, $\mu_2$, resulting in a different scattering form factors for these two types of ions. Within this model, the CDW resonant diffraction peak intensity is proportional to $|F_1 - F_2|^2$, where $F_1$ and $F_2$ are the scattering form factors for the ions at type-1 and type-2 rare earth sites, respectively. The imaginary part of the form factor is the absorption and thus the form factors can be reconstructed from the absorption spectrum by utilizing Kramers-Kronig transformation.

The absorption spectrum for a type-2 site $\mu_2$ is just that of a spherically symmetric $R^{3+}$ ion. This spectrum is calculated in the usual way as a ground state averaged over each of the degenerate Hund's rule ground states; in spherical symmetry contributions to the XAS from the next highest electronic initial state are negligible at all experimentally relevant temperatures due to the Boltzmann factor. For the type-1 sites, multiplet energy levels are given by:
\begin{equation}\label{eq:energy}
E_J \to E_{J,M}=E_J+\frac{\Delta}{J}|M|\;,
\end{equation}
where $J(J+1)$ is the eigenvalue of the total angular momentum and $M$ the eigenvalue of the z-component of the total angular momentum. The parameter $\Delta$ has the dimension of energy and has been introduced to account for the effect of the CDW on the type-1 sites. We have divided $\Delta$ by J in Eq.~(\ref{eq:energy}) so that $\Delta$ represents the total splitting over the entire multiplet. The magnitude of $\Delta$ will be seen to determine the temperature dependence of the REXS peak, and can thus be estimated from the experimental data shown in Fig. \ref{Fig3:T_dep_TbTe3}(c). The total XAS and resonant X-ray diffraction peak intensity containing these two types of ions can be obtained and compared to our experiment. Importantly, as $\Delta$ is comparable to the energy scale of the measurement temperature, Boltzmann statistics were also included to account for the thermal population of the split $4f$ multiplets.

\begin{figure*}[t]
\includegraphics [clip, width=6.0 in]{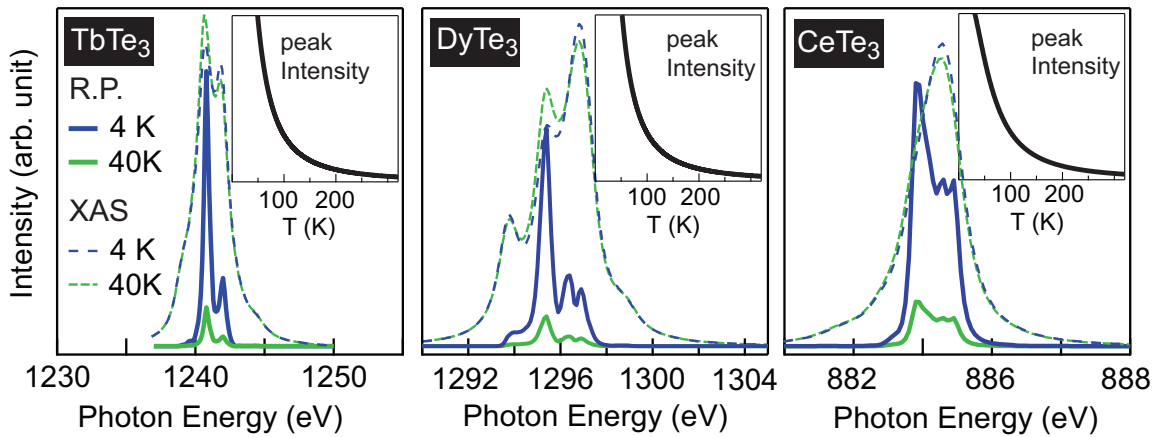}
\centering
\caption{\label{Fig6:Calculation_XAS_RP} (Color online) The calculated resonant profile (R. P.) of the CDW diffraction peak and X-ray absorption spectrum (XAS) for TbTe$_3$, DyTe$_3$, and CeTe$_3$ near the \emph{M5} edge of rare earth elements at two representative temperatures, 4 K and 40 K. The calculated temperature dependence of the CDW diffraction peak intensity taken at the photon energy equal to the maximum of the R.P. is plotted in the insets.
}
\end{figure*}

To calculate the XAS, we first consider the effect of a non-zero $\Delta$ on the XAS. In general, the XAS is given by an average over initial states $\ket{I}$ and a sum over final states $\ket{F}$:
\begin{equation}\label{eq:general}
\mu(\omega)=\sum_I \frac{e^{-E_I/T}}{Z} \sum_{F}{|\bra{F}\epsilon\cdot {\bf r}\ket{I}|}^2\delta(\omega+E_I-E_F)\;,
\end{equation}
where we have set the Boltzmann constant $k=1$, $Z=\sum_I e^{-E_I/T}$ is the partition function, $T$ is the temperature, $\omega$ is the incident photon energy, $\epsilon$ is the polarization, and ${\bf r}$ is the sum of the position operators of the electrons on the ion. Eq.~\ref{eq:general} is a general formula for dipole absorption, however, we may simplify the formula if we realize that the $\Delta = 0$ Hund's rule ground state energy is well-separated from the next highest excited state and thus the average over initial states can be restricted to
an average over the $\Delta=0$ ground state manifold for all experimentally relevant temperatures.
For non-zero $\Delta$ the energy depends on both $J$ and $|M|$ and we have
\begin{equation}\label{eq:lessgeneral}
\begin{split}
&\mu(\omega) =\sum_{M=-J_0}^{J_0} \frac{e^{-\Delta|M|/T}}{Z_{J_0}} \times \\ &\sum_{J_F,M_F,x_F}{|\bra{J_F,M_F,x_F}
\epsilon\cdot {\bf r} \ket{J_0,M,x_0}|}^2 \times \\
&\delta_\Gamma(\omega+E_{J_0}-E_{J_F})\;,
\end{split}
\end{equation}
where $Z_{J_0}=\sum_{M=-J_0}^{J_0}e^{-\Delta|M|/T}$, and $x_F$ and $x_0$ are any additional labels needed to specify the state.
In Eq.~(\ref{eq:lessgeneral}) we have replaced the delta function with a broadened Lorentizian $\delta_\Gamma$ where the broadening is due to the lifetime of the core-hole. Additionally, because we assume that the width due to the core hole lifetime is much greater than $\Delta$ we have ignored the $M$ dependence within the argument of the Lorentzian.\cite{Goedkoop88} However, we make no assumption about the relative size of $T$ and $\Delta$, thus the $|M|$ dependence of the Boltzmann factor must be retained.
Eq.~(\ref{eq:lessgeneral}) can be written in terms of reduced matrix elements and Wigner Three-J symbols as:
\begin{equation}\label{eq:lessgeneral2}
\begin{split}
&\sum_{J_f,M_f,x_f}\!\!{|\!\bra{J_f,x_f}\!\!|r|\!\!\ket{J_0,x_0}\!|}^2 \times \\
&\left(
\frac{1}{Z_{J_0}}\sum_{M=-J_0}^{J_0} e^{-\Delta |M|/T}
{|\threej{J_f}{1}{J}{-M_f}{q}{M}\epsilon_q^*|}^2
\right) \times \\
&\delta_{\Gamma}(E_{J_0}+\omega-E_{J_f})\;,
\end{split}
\end{equation}
where the quantity in the large parenthesis gives a temperature dependent weighing factor for the different possible
final state $J_f$ values ($J_0-1$, $J_0$, $J_0+1$). Interestingly, this weighing factor can be written in the convenient form\cite{Thole85}:
\begin{equation}
\begin{split}
&\left(
\frac{1}{Z_{J_0}}\sum_{M=-J_0}^{J_0} e^{-\Delta |M|/T}
{|\threej{J_f}{1}{J_0}{-M_f}{q}{M}\epsilon_q^*|}^2
\right) \\
&=
a_{J_0,J_f}+b_{J_0,J_f}<{M^2}>(\Delta/T)
\end{split}
\end{equation}

where the right hand side follows from the fact that
(for $J_f$ in the allowed range)
the squared Wigner Three-J symbol is in fact a quadratic polynomial in $M$.

Thus we have a final expression for the temperature dependent XAS
from a single ion of type $i$:
\begin{equation}
\begin{split}
&\mu_i(\omega)=\sum_{J_f=J_0-1}^{J_0+1}\mu_{J_0, J_f}(\omega) \times \\
&\left(
a_{J_0,J_f}+b_{J_0,J_f}<M^2>((2-i)\Delta/T)
\right)
\end{split}
\end{equation}
where $i=1$ or $2$, $\Delta$ is the fixed symmetry breaking energy due to the frozen CDW distortion, and where $\mu_{J_0,J_f}$ is the partial absorption for transitions to final states with angular momentum $J_f$. The partial absorption can be obtained from well-known computer software such as Cowan's code\cite{CowansCode}.
Note that for $i=2$ the argument of the average square moment is $(2-2)\Delta/T=0$ and the XAS is just equal to the spherically symmetric spectrum.

Each partial absorption $\mu_{J_0,J_f}$ can be Kramers-Kronig transformed to obtain the partial REXS scattering form factor as
$$
F_{J_0,J_f}=\frac{1}{\pi}\int dz \frac{1}{z-(\omega+i\delta)}\mu_{J_0,J_f}(\omega)\;,
$$
and
$$
F_i=\sum_{J_f=J_0-1}^{J_0+1}F_{J_0,J_f}
\left(
a_{J_0,J_f}+b_{J_0,J_f}<M^2>
\right)\;.
$$

In Fig.~\ref{Fig6:Calculation_XAS_RP}, we plot the calculated XAS and resonant profile of the resonant diffraction peak for the TbTe$_3$, DyTe$_3$, and CeTe$_3$ near the \emph{M5}- edge of rare earth elements. For temperature dependence calculation, we ignore the temperature dependence of CDW's order parameter; such that the calculated temperature dependent spectrum reflect solely the effect due to the thermal occupancy of $4f$-multiplet. The calculated XAS reproduce all major features observed in the experiments (see Fig. \ref{Fig2:RP_TbTe3} and Fig. \ref{Fig4:OtherRTe3}), suggesting that this model captures the essential $4f$ multiplet physics involved in the $M_5$ absorption edge. Importantly, the calculated XAS exhibit a weak temperature dependence, as observed in the experiment. Regarding the resonant profile of CDW diffraction peak intensity, the calculated resonant profiles also capture the major characteristic feature of the experimental data (Fig. \ref{Fig2:RP_TbTe3} and Fig. \ref{Fig4:OtherRTe3}), exhibiting a distinct shape than the XAS. Furthermore, unlike the XAS, the intensity of the calculated resonant profile enhances dramatically at low temperatures, while the overall lineshape of the resonant profile is still preserved. Our calculation shows that the resonant diffraction peak intensity increases exponentially, as observed experimentally. This exponential-like temperature dependence is due to the fact that $\Delta$ is of the same order as the measurement temperature; as a consequence, the thermal occupancy of the split $4f$ multiplet, which is governed by Boltzmann factor, dominates the temperature dependence of the CDW resonant diffraction peak intensity.

Also, consistent with the experimental observations, the REXS intensity for both LaTe$_3$ and GdTe$_3$ is identically zero in theory. This is obviously the case for LaTe$_3$ which has a non-degenerate $J=0$ ground state and can not be further split by the symmetry breaking field. The GdTe$_3$ ground state is spatially symmetric $L=0$ but degenerate due to the spin $S=7/2$. However, a spatial perturbation such as that due to the CDW field can not split the spin degeneracy and thus there can not be a REXS signal for the GdTe$_3$ system either. This can be understood as follows. A general CDW perturbation $\delta H$ which acts only on spatial variables can be written in second quantization as
$$
\delta H =\sum_{a,b,\sigma}c^\dagger_{a,\sigma}c_{b,\sigma}v_{a,b}\;,
$$
where the $v_{a,b}$ are numbers, the indices $a$ and $b$ refer to a complete set of spatial basis functions, and the index $\sigma$ labels the spin. Using the explicit form given above it is straightforward to show
$$
[\delta H, \vec S]=0\;.
$$
In order for the perturbation not to spilt the degeneracies of any atomic multiplet states, we might try to show that for GdTe$_3$
$$
[\delta H,\vec J]\stackrel{?}{=}0\;.
$$
This is not generally true, however, we only require a weaker condition: $\delta H$ commute with $\vec J$ within the ground state manifold. But, within the ground state manifold we have already seen that we may set $L=0$. Thus,
$$
[\delta H, \vec J]=[\delta H,\vec L+\vec S]\to[\delta H,\vec S]=0\;.
$$
Because the perturbation commutes with $J$ in the ground state manifold the total Hamiltonian commutes with $J$ which thus remains a good quantum number.

\section{Discussion}

 Usually, the CEF parameters are determined by inelastic neutron scattering, or, by optical spectroscopy -- two techniques for which the associated energy scale is clearly well suited. However, these two techniques measure spatially averaged CEF parameter, not able to distinguish a periodic modulated CEF effect, like the present cases. Here, using resonant X-ray diffraction, the 4$f$ multiplet splitting parameter, $\Delta$, can be estimated. We fit the experimental temperature dependence data to the theoretical model, yielding value of $\Delta$ for Tb, Dy, and Ce to be approximately 4 meV, 2 meV, and 20 meV, respectively.  Therefore, when compared to typical electronic energy scales, the $4f$ multiplet splitting is small and thus difficult to resolve for spectroscopies which measure electronic structure such as XAS. The REXS, however, is sensitive to this small energy scale due to the quantum interference which results in a spectrum proportional to $|F_1-F_2|$.

We also remark that the values of $\Delta$ fitted from our data are on the order of the expected change in $4f$-$2p$ hybridization energy due to the CDW distortion. The hybridization energy between the rare earth $4f$ electrons and Te $2p$ electrons is estimated to be\cite{Harrison} approximately $0.3$ eV. This energy varies with distance as the inverse fifth power of the distance between ions\cite{Harrison}. Thus the magnitude of the percent change in hybridization energy is five times the percentage change in bond distance associated with the CDW distortion.  According to non-resonant X-ray diffraction{\cite{Nancy08}}, the change of the bond length is on the order of $0.3$ percent in TbTe$_3$, corresponding to a $\Delta$ approximately 5 meV, or 60K. This is consistent with the value of $\Delta$ estimated from the experimental data, and a similar order of magnitude to the other estimated values from experiments. A more detailed theory of the tri-telluride materials could also include further site-types for the rare-earth ions and could attempt an {\it ab initio} calculation of spatially-dependent splitting parameters at each site. However, we do not pursue this line here as we are interested only in the qualitative effect of the CDW on the rare-earth elastic scattering.


In summary, the periodic 4$f$ multiplet splitting of rare earth ions induced by a charge density wave (CDW) state in the family of rare earth tri-telluride was detected and measured by resonant soft X-ray diffraction near the rare earth $M_5$ absorption edge. The diffraction peak intensity exhibits an exponential increase at low temperatures, which is due to the variation of the thermal occupancy of rare earth 4$f$ multiplet. Our results also demonstrate that the resonant peak intensity does not necessarily simply represent the order parameters of CDW state; thermal averaging of the ground states and intermediate states can also contribute to the measured temperature dependence of the diffraction peak, when their band width are comparable to the measurement temperatures.

\section{Acknowledgement}
This work is supported by the U.S. Department of Energy, Office of Basic Energy Sciences, Division of Material Sciences and Engineering under Contract No. DE-AC02-76SF00515, SLAC National Accelerator Laboratory (SLAC), Stanford Institute for Materials and Energy Sciences. The Advanced Light Source is supported by the Director, Office of Sciences, Office of Basic Energy Sciences, of the U.S. Department of Energy under Contract No. DE-AC02-05CH11231.


\begin{thebibliography}{99}
\bibitem{Hill95}
J. P. Hill, G. Helgesen, and D. Gibbs, Phys. Rev. B \textbf{51}, 10336 (1995).

\bibitem{Abbamonte02}
P. Abbamonte \emph{et al.}, Science \textbf{297}, 581 (2002).

\bibitem{Wilkins03}
S. B. Wilkins \emph{et al.}, Phys. Rev. Lett \textbf{91}, 167205 (2003).

\bibitem{Thomas04}
K. J. Thomas \emph{et al.}, Phys. Rev. Lett. \textbf{92}, 237204 (2004).

\bibitem{Abbamonte04}
P. Abbamonte \emph{et al.}, Nature \textbf{431}, 1078 (2004).

\bibitem{Dhesi92}
S. S. Dhesi \emph{et al.}, Phys. Rev. Lett. \textbf{92}, 056403 (2004).

\bibitem{Christian05}
C. Sch\"{u}{\ss}ler-Langeheine \emph{et al}. Phys. Rev. Lett. \textbf{95}, 156402 (2005).

\bibitem{Nancy06}
N. Ru and I. R. Fisher, Phys. Rev. B \textbf{73}, 033101 (2006).

\bibitem{Nancy08}
N. Ru, C. L. Condron, G. Y. Margulis, K. Y. Shin, J. Laverock, S. B. Dugdale, M. F. Toney, and I. R. Fisher, Phys. Rev. B \textbf{77}, 035114 (2008).

\bibitem{Banerjee}
A.Banerjee and T. Rosenbaum, private communication.

\bibitem{Abbamonte06}
P. Abbamonte, Phys. Rev. B \textbf{74}, 195113 (2006).

\bibitem{Gweon98}
G. H. Gweon \emph{et al.}, Phys. Rev. Lett. \textbf{81}, 886 (1998).

\bibitem{Brouet04}
V. Brouet \emph{et al.}, Phys. Rev. Lett. \textbf{93}, Phys. Rev. Lett. \textbf{93}, 126405
 (2004).

 \bibitem{Brouet08}
 V. Brouet, W. L. Yang, X. J. Zhou, Z. Hussain, N. Ru, K. Y. Shin, I. R. Fisher, Z. X. Shen, Phys. Rev. B \textbf{77}, 235104 (2008).

 \bibitem{Moore08}
 R. G. Moore, V. Brouet, R. He, D. H. Lu, N. Ru, J.-H. Chu, I. R. Fisher, and Z.-X. Shen, Phys. Rev. B \textbf{81}, 073102 (2010).

\bibitem{Goedkoop88}
J. B. Goedkoop, B. T. Thole, G. van der Laan, and G. A. Sawatzky, F. M. F. de Groot and J. C. Fuggle, Phys. Rev. B \textbf{37}, 2086 (1988)

\bibitem{Thole85}
B. T. Thole, G. van der Laan, and G. A. Sawatzky, Phys. Rev. Lett. \textbf{55}, 2086 (1985).

\bibitem{CowansCode}
 R. D. Cowan, "The Theory of Atomic Structure and Spectra", University of California Press, 1981.
\bibitem{Harrison}
Walter A. Harrison, \emph{Elementary electronic structures} (World Scientific Publishing 2004).
\end{thebibliography}
\end{document}